\documentclass[11pt]{article}
\pdfoutput=1 

\usepackage{jheppub} 
\usepackage{graphicx}

\usepackage[T1]{fontenc} 
\usepackage{amsmath}

\newcommand\fft[2]{\frac{#1}{#2}}
\newcommand\ft[2]{{\textstyle\frac{#1}{#2}}}
\newcommand\nn{\nonumber}

\title{Seeing bulk perturbations in Lifshitz holography}

\preprint{MCTP-17-04}

\author{James T. Liu and}
\author{Pranav V. Rao}

\affiliation{Michigan Center for Theoretical Physics, Randall Laboratory of Physics\\
The University of Michigan, Ann Arbor, MI 48109-1040, USA }
\emailAdd{jimliu@umich.edu}
\emailAdd{pvr@umich.edu}

\abstract{We introduce a perturbation $h_{\mu\nu}$ onto a background Lifshitz spacetime and examine some of its consequences.  In particular, we consider a radially localized perturbation and compute the resulting holographic Green's function to linearized order.  At leading order, the Lifshitz Green's function demonstrates suppression of spectral weight at low frequencies, and this feature allows bulk perturbations in the IR to be partially hidden from local boundary probes.}

\begin{document} 
\maketitle
\flushbottom

\section{Introduction}
\label{sec:intro}

The question: `can holography be applied to condensed matter physics?' excites the many who intuit the answer is positive. Still, making connections with materials and garnering novel insight to inform both sides of the gauge-gravity duality is an ongoing effort. The nature of this duality leads to the focus on regions near quantum critical points, where important properties of the system are governed by scale invariance.  In the relativistic case, this is generally extended to full conformal invariance, while for condensed matter systems, the theory at the quantum critical point may be described by Lifshitz or Schr\"odinger symmetry with possible hyperscaling violation.

While much initial work on non-relativistic holography have been limited to systems with Euclidean or Galilean symmetry, realistic condensed matter systems are generally categorized as having a space or point group discrete symmetry --- a lattice --- that breaks these continuous symmetries. This has led to the introduction of holographic models that explicitly break translational invariance \cite{Horowitz:2012ky, Horowitz:2012gs,Donos:2012js,Vegh:2013sk,Horowitz:2013jaa,Chesler:2013qla,Ling:2013nxa,Blake:2013owa,Donos:2013eha,Andrade:2013gsa}.  Moreover, the addition of impurities, or more generally disorder \cite{Hartnoll:2007ih,Fujita:2008rs,Hartnoll:2008hs,Ryu:2011vq,Harrison:2011fs,Arean:2013mta,Davison:2013txa,Hartnoll:2014cua,Lucas:2014zea,Adams:2011rj} has extended the range of phenomena that can be addressed via non-relativistic holography.

What the addition of a lattice or random disorder does is to modify the bulk geometry away from the simple case of, e.g., a pure Lifshitz background.  While, in general, such modifications would be spread out along the radial (i.e.~holographic) direction, it is also possible to consider the effect of localized perturbations on the system.  Moreover, even if the disorder is not radially localized, so long as it is sufficiently weak, it can be treated using linearized perturbation theory, and hence can be viewed as a superposition of localized perturbations.

In contrast with the relativistic case, where bulk perturbations are causally connected to the boundary and can directly influence boundary dynamics, in non-relativistic holography such perturbations can be `hidden' from the boundary.  In particular, under certain conditions, modifications to the IR geometry or the horizon boundary conditions will leave only an exponentially small imprint on local observables in the boundary theory  \cite{Keeler:2013msa, Keeler:2014lia}.  One consequence of this is that the smearing function \cite{Bena:1999jv,Hamilton:2005ju,Hamilton:2006az} does not exist for a Lifshitz bulk, and another is the suppression of spectral weight in the low-frequency limit \cite{Faulkner:2010tq,Hartnoll:2012wm,Keeler:2013msa}.

In this paper, we explore in more detail the effect of a perturbation in the Lifshitz bulk on the holographic scalar Green's function $G(\omega,\vec k)$ of the boundary theory and quantify the range of frequency and momenta for which it becomes insensitive to the perturbation.  We will focus on a Lifshitz bulk with unperturbed metric
\begin{equation}
\label{eq:lifmet:1}
ds_{(0)}^2 = g_{\mu\nu}^{(0)}dx^\mu dx^\nu
=- \left(\frac{L}{r} \right)^{2z} dt^2 + \left(\frac{L}{r}\right)^2 (d\vec{x}_d^2+ dr^2),
\end{equation}
and add to it a perturbation of the form
\begin{equation}
\label{eq:lifmet:3}
g_{\mu \nu}  = g^{(0)}_{\mu \nu} + \kappa \, h_{\mu \nu}.
\end{equation}

At linearized order in the metric perturbation, the bulk scalar equation, $(\square-m^2)\phi=0$, takes the form
\begin{equation}
(\square_0-m^2)\phi=\kappa\left[-\ft12g_0^{\mu\nu}\partial_\mu h\partial_\nu+\nabla^0_\mu h^{\mu\nu}\partial_\nu+h^{\mu\nu}\nabla_\mu^0\partial_\nu\right]\phi,
\label{eq:lineom}
\end{equation}
where $h=h^\mu_\mu$.  This equation can be solved perturbatively for $\phi$.  In particular, we let $\phi=\phi_0+\kappa\phi_1$, where $\phi_0$ solves the unperturbed equation $(\square_0-m^2)\phi_0=0$.  The perturbed field $\phi_1$, then solves (\ref{eq:lineom}) with $\phi_0$ substituted on the right-hand side.  As a result, we need to solve the bulk scalar equation of motion with a source term.  This may be accomplished by obtaining the bulk scalar Green's function
\begin{equation}
(\square_0-m^2)G(t,\vec x,r;t_0,\vec x_0,r_0)=\fft1{\sqrt{-g^{(0)}}}\delta(t-t_0)\delta^d(\vec x-\vec x_0)\delta(r-r_0).
\label{eq:bulkG}
\end{equation}
In fact, the Green's function itself represents the response to a fully localized perturbation in the bulk.  Thus we can extract the effects of bulk perturbations by direct examination of the Green's function solution, and this is the approach that we take below.

We will primarily work with $z=2$ Lifshitz where analytical results are possible, but will extend our results to general $z>1$ using the WKB approximation.  In the following section, we will compute the perturbed holographic two-point function, and in
section~\ref{sec:expsup} we will examine its properties, and in particular address the question of when the bulk perturbation is visible or hidden from the boundary.  We then conclude in section~\ref{sec:con}.

\section{Holographic two-point functions}

We briefly review the linear-response prescription for the holographic Green's functions for general $z$, and then focus on the $z=2$ Lifshitz case, where an analytic solution is available.  For the unperturbed metric \eqref{eq:lifmet:1}, the bulk scalar equation $(\square_0-m^2)\phi=0$ reduces to
\begin{equation}\label{eq:kgR0}
\left[ \partial_r^2 - \frac{z+d-1}{r} \partial_r - \vec k^2 + \left(\frac{r}{L}\right)^{2(z-1)} \omega^2 - \frac{(mL)^2}{r^2}\right]f(r) = 0,
\end{equation}
where we have taken $\phi(t,\vec x,r)=f(r)e^{i(\vec k\cdot\vec x-\omega t)}$.  At the boundary, $r \rightarrow 0$, the asymptotic behavior is given by a  sum of power laws,
\begin{equation}\label{eq:bdry}
f \sim B r^{\Delta_+} + A r^{\Delta_-},
\end{equation}
where
\begin{equation}\label{eq:dimn}
\Delta_{\pm} = \frac{d+z}{2} \pm\nu,\qquad\nu= \sqrt{ (mL)^2 + \left( \frac{d+z}{2} \right)^2}.
\end{equation}
Here $B$ and $A$ are the coefficients of the normalizable and non-normalizable modes, respectively.
For $z>1$, the near-horizon ($r\to\infty$) behavior is oscillatory, and takes the form
\begin{equation}\label{eq:hrzn}
f \sim \left(\frac{r}{L} \right)^{d/2}\left\lbrace a \exp{\left(i\frac{\omega L}{z} \left(\frac{r}{L}\right)^z\right)} + b  \exp{\left(-i\frac{\omega L}{z} \left(\frac{r}{L}\right)^z\right)}\right\rbrace.
\end{equation}
The holographic retarded Green's function can then be computed as the ratio of the normalizable mode (response) over the non-normalizable mode (source), assuming in-falling boundary conditions at the horizon ($b=0$), consistent with causality requirements
\begin{equation}\label{eq:grfn}
G_R(\omega, k) = \frac{B}{A} \biggr |_{b=0}
\end{equation}
This requires connecting the horizon behavior, specified by $a$ and $b$, to the boundary behavior, given by $A$ and $B$.

\subsection{The zeroth order solution}

In general, the scalar equation (\ref{eq:kgR0}) is not solvable analytically and must be treated via WKB or other approximate methods.  However, for $z=1$ (AdS) and $z=2$ (Lifshitz), the scalar equation is exactly solvable in terms of Bessel and Whittaker functions, respectively.  In any case, we find it convenient to rewrite the second order equation into Schr\"odinger form and to transform to dimensionless variables.  To do so, we take
\begin{equation}
f=r^{d/2}\psi,\qquad \zeta=\fft{(\omega L)(r/L)^z}z.
\label{eq:rzeta}
\end{equation}
The resulting equation is then
\begin{equation}\label{eq:schrodinger}
-\fft{d^2}{d\zeta^2}\psi(\zeta)+U(\zeta)\psi(\zeta)=0,
\end{equation}
where
\begin{equation}\label{eq:pot}
U(\zeta)=\fft{(\nu/z)^2-1/4}{\zeta^2}+\fft{\alpha}{\zeta^{2(1-1/z)}}-1,\qquad
\alpha=\left(\fft{kL/z}{(\omega L/z)^{1/z}}\right)^2.
\end{equation}

For general $z$, we may consider a WKB approximation.  Although this approximation must be treated with care, it can be used to extract the imaginary part of the Green's function (i.e.~the spectral function).  Taking infalling boundary conditions for the zeroth order solution gives
\begin{equation}
\psi_+(\zeta)=\begin{cases}\fft{e^{-i\pi/4}}{(\hat U(\zeta))^{1/4}}\left[e^{S(\zeta,\zeta_*)}+\fft{i}2e^{-S(\zeta,\zeta_*)}\right],&\zeta<\zeta_*;\\
\fft1{(-\hat U(\zeta))^{1/4}}e^{i\Phi(\zeta_*,\zeta)},&\zeta>\zeta_*,\end{cases}
\label{eq:WKB+}
\end{equation}
where
\begin{equation}
\label{eq:tunneling}
S(\zeta,\zeta_*)=\int_\zeta^{\zeta_*}\sqrt{\hat U(\zeta')}d\zeta',\qquad
\Phi(\zeta_*,\zeta)=\int_{\zeta_*}^\zeta\sqrt{-\hat U(\zeta')}d\zeta',
\end{equation}
and $\zeta_*$ is the classical turning point, $\hat U(\zeta_*)=0$.  Here $\hat U(\zeta)=U(\zeta)+1/4\zeta^2$ is the shifted potential appropriate to the WKB approximation for an inverse-squared potential.  For later construction of the bulk scalar Green's function, we also note that WKB solution that is normalizable at the boundary is given by
\begin{equation}
\psi_n(\zeta)=\begin{cases}\fft1{(\hat U(\zeta))^{1/4}}e^{2S(\zeta,\zeta_*)},&\zeta<\zeta_*;\\
\fft2{(-\hat U(\zeta))^{1/4}}\cos(\Phi(\zeta_*,\zeta)-\ft\pi4),&\zeta>\zeta_*.\end{cases}
\end{equation}

The holographic Green's function can be extracted by the boundary behavior of $\psi_+$, which has both a normalizable $\sim e^{-S}$ and a non-normalizable $\sim e^S$ component.  Transforming back to the radial coordinate $r$ and extracting the $A$ and $B$ coefficients according to (\ref{eq:bdry}) then gives the WKB approximation to the spectral function
\begin{equation}
\chi(\omega,k)=2\Im G_R(\omega,k)=L^{-2\nu}\left(\fft{\omega L}z\right)^{2\nu/z}\lim_{\epsilon\to0}
\epsilon^{-2\nu/z}e^{-2S(\epsilon,\zeta_*)}.
\end{equation}
The behavior of the spectral function depends on the location of the turning point $\zeta_*$.  For $\omega L/\nu>(kL/\nu)^z$, the turning point lies in the $1/\zeta^2$ region of the effective Schr\"odinger potential, and the spectral function has a power-law behavior, $\chi\sim \omega^{2\nu/z}$.  For smaller $\omega$, the spectral weight becomes exponentially suppressed, and a matched asymptotic expansion gives \cite{Faulkner:2010tq}
\begin{equation}
\chi\approx\left(\fft{ek}{2\nu}\right)^{2\nu}\exp\left[-\fft{\sqrt\pi}z\fft{\Gamma(\fft1{2(z-1)})}{\Gamma(\fft{z}{2(z-1)})}\left(\fft{(kL)^z}{\omega L}\right)^{\fft1{z-1}}\right].
\label{eq:chiexp}
\end{equation}

\subsubsection{The $z=2$ case}

When $z=2$, the WKB approximation is not needed since the scalar equation can be solved in terms of Whittaker functions.  The infalling and normalizable solutions are given by
\begin{equation}
\psi_+(\zeta) = W_{-i \alpha/2, \nu/2} (-2i\zeta) ,\qquad\psi_n= M_{-i\alpha/2,\nu/2}(-2i\zeta),
\label{eq:whit}
\end{equation}
where $\alpha = {k^2L}/{2 \omega}$.  For the retarded Green's function, we take the infalling solution $\psi_+$ and make use of the small argument expansion \cite{NIST:DLMF}
\begin{equation}
\label{eq:boundaryW}
W_{-i \alpha/2, \nu/2} (-2i\zeta) \bigg|_{\zeta \ll 1} \approx\frac{\Gamma(\nu)}{\Gamma(\fft{1+\nu + i\alpha}2)}(-2i\zeta)^{(1-\nu)/2} +\frac{\Gamma(-\nu)}{\Gamma(\fft{1 - \nu+ i\alpha}2)}(-2i\zeta)^{(1+\nu)/2},
\end{equation}
to obtain \cite{Kachru:2008yh}
\begin{equation}
G_R(\omega,k) =  \left(\fft{-i \omega}L\right)^{\nu}\frac{\Gamma(-\nu)}{\Gamma(\nu)}\frac{\Gamma(\fft{1+\nu + i \alpha}2)}{ \Gamma(\fft{1-\nu + i \alpha}2)}.
\label{eq:GRz2}
\end{equation}
Expanding this for large $\alpha$ shows that there is exponential suppression in the spectral function on the order of $e^{-\pi \alpha}$.  We now consider the perturbed case and seek to show that such suppression still exists in the first-order shift in the holographic Green's function in this large $\alpha$ limit.

\subsection{The first order solution}

In order to obtain the first order solution due to a bulk metric perturbation, we first construct the bulk scalar Green's function (\ref{eq:bulkG}).  Taking the unperturbed metric to be of the form (\ref{eq:lifmet:1}) and working in momentum space gives
\begin{equation}
G(t,\vec x,r;t',\vec x',r')=\int\fft{d\omega}{2\pi}\fft{d^dk}{(2\pi)^d}g_{\omega,k}(r,r')
e^{i(\vec k\cdot(\vec x-\vec x')-\omega(t-t'))},
\end{equation}
where $g_{\omega,k}(r,r')$ satisfies the bulk equation (\ref{eq:kgR0}) with a delta-function source
$(r/L)^{z+d-1}\delta(r-r')$ on the right-hand side.  Transforming to the dimensionless $\zeta$ coordinate given in (\ref{eq:rzeta}) and working out the jump condition at the delta function then gives the bulk scalar Green's function
\begin{equation}
g_{\omega,k}(r,r')=\fft1\omega\left(\fft{rr'}{L^2}\right)^{d/2}\fft{\psi_n(\zeta_<)\psi_+(\zeta_>)}W,
\end{equation}
where $W=\psi_n(d\psi_+/d\zeta)-(d\psi_n/d\zeta)\psi_+$ is the Wronskian.  Here $\zeta_<=\min(\zeta,\zeta')$ and $\zeta_>=\max(\zeta,\zeta')$.  Note that we have taken the boundary conditions to be infalling and the horizon and normalizable at the boundary.

Given this bulk scalar Green's function, the first order solution to (\ref{eq:lineom}) then takes the form
\begin{equation}
\phi(\omega,\vec k,r)=\phi_0(\omega,\vec k,r)+\kappa\int_0^\infty\left(\fft{L}{r'}\right)^{z+d+1}dr'\, g_{\omega,k}(r,r')\rho(\omega,\vec k,r').
\end{equation}
The source $\rho(\omega,\vec k,r')$ can be obtained from the right-hand side of (\ref{eq:lineom}).  While the general expression is rather lengthy and not all that enlightening, it takes a more concise form for traceless metric perturbations, $h=0$.  In this case, we find
\begin{equation}
\rho(\omega,\vec k,r')=-\int\fft{d^{d+1}k'}{(2\pi)^{d+1}}
k_\mu k'_\nu h^{\mu\nu}(k-k',r')\phi_0(\omega',\vec k',r'),
\end{equation}
which is simply the convolution of the zeroth order solution with the traceless metric perturbation.  Here we are working in the gauge $h_{\mu\,r}=h_{rr}=0$, and we have taken $k^\mu=(\omega,\vec k)$ to streamline the notation.

Since we are interested in extracting the retarded Green's function from the boundary behavior of $\phi(\omega,\vec k,r)$, we take the limit $r\to0$.  Substituting in the zeroth order solution $\phi_0=r^{d/2}\psi_+$ and transforming to the dimensionless $\zeta$ variable then gives
\begin{align}
\phi(\omega,\vec k,r)\bigg|_{r\to0}&=r^{d/2}\Biggl[
\psi_+(k,\zeta)\nn\\
&\quad-\kappa\fft{L^2}{z^2W}\psi_n(k,\zeta)\int_{\zeta_c}^\infty\fft{d\zeta'}{\zeta'^2}\int\fft{d^{d+1}k'}{(2\pi)^{d+1}}k_\mu k'_\nu h^{\mu\nu}(k-k',\zeta')\psi_+(k,\zeta')\psi_+(k',\zeta')\Biggr],
\label{eq:phi01}
\end{align}
where we have restored the $(\omega,\vec k)$ dependence of the infalling, $\psi_+$, and normalizable, $\psi_n$, wavefunctions.  Here we have assumed the metric perturbation is contained in the bulk, so that $h_{\mu\nu}(k,\zeta')$ vanishes for $\zeta<\zeta_c$ for a fixed $\zeta_c$.

The convolution in (\ref{eq:phi01}) arises because momentum is not conserved for general metric perturbations.  However, for a translationally invariant perturbation $h_{\mu\nu}(r)$, the above simplifies to
\begin{equation}
\phi(\omega,\vec k,r)\bigg|_{r\to0}=r^{d/2}\left[
\psi_+(k,\zeta)-\kappa\fft{L^2}{z^2W}\psi_n(k,\zeta)\int_{\zeta_c}^\infty\fft{d\zeta'}{\zeta'^2}k_\mu k_\nu h^{\mu\nu}(\zeta')\left(\psi_+(k,\zeta')\right)^2\right].
\end{equation}
In order to extract the retarded Green's function, we parameterize the boundary behavior of $\psi_+$ and $\psi_n$ by
\begin{align}
\psi_+(\zeta)&\sim \hat A_+\zeta^{1/2-\nu/z}+\hat B_+\zeta^{1/2+\nu/z},\nn\\
\psi_n(\zeta)&\sim \hat B_n\zeta^{1/2+\nu/z}.
\end{align}
This allows us to compute the Wronskian, and we find $W=-2(\nu/z)\hat A_+\hat B_n$.  Using (\ref{eq:grfn}), we finally obtain
\begin{equation}
G_R(\omega,k)=L^{-2\nu}\left(\fft{\omega L}z\right)^{2\nu/z}\left[
\fft{\hat B_+}{\hat A_+}+\kappa\fft{L^2}{2z\nu}\int_{\zeta_c}^\infty\fft{d\zeta'}{\zeta'^2}k_\mu k_\nu h^{\mu\nu}(\zeta')\left(\fft{\psi_+(\zeta')}{\hat A_+}\right)^2\right].
\label{eq:GR01}
\end{equation}

\subsubsection{A radially localized perturbation}

The expression (\ref{eq:GR01}) gives the first order shift in the retarded Green's function for translationally
invariant perturbations to the bulk metric in terms of an integral over the radial profile of the perturbation
multiplied by the square of the normalized infalling wavefunction $\psi_+(\zeta)$.  The reason for the
square is that the boundary source has to propagate into the bulk, then scatter off the metric perturbation,
and finally propagate back out to the boundary.

To get a general idea of the behavior of the $G_R(\omega,k)$ and its sensitivity to bulk perturbations,
we now consider a radially localized metric fluctuation
\begin{equation}
h_{\mu\nu}=\left(\fft{L}r\right)^2\mathcal A_{ij}\delta((r-r_0)/L),
\end{equation}
where $\mathcal A_{ij}$ is constant and has vanishing trace.  Substituting this into (\ref{eq:GR01}) then
gives
\begin{equation}
G_R(\omega,k)=L^{-2\nu}\left(\fft{\omega L}z\right)^{2\nu/z}\left[\fft{\hat B_+}{\hat A_+}
+\kappa\mathcal A_{ij}k^ik^jL^2\fft{z}{2\nu\omega L}\left(\fft{L}{r_0}\right)^{z-1}
\left(\fft{\psi_+(\zeta_0)}{\hat A_+}\right)^2\right],
\end{equation}
where $\zeta_0=(\omega L)(r_0/L)^z/z$ from (\ref{eq:rzeta}).
Since scale invariance is broken by the perturbation at bulk radius $r_0$, the Green's function now depends non-trivially
on $\omega$ and $k$.  For fixed $r_0$, we can introduce the scaled quantities
\begin{equation}
\hat\omega=(\omega L)\left(\fft{r_0}L\right)^z,\qquad\hat k=kr_0.
\end{equation}
In this case, we have
\begin{align}
G_R(\hat\omega,\hat k) &= G^{(0)}_R(\hat\omega,\hat k) + \kappa G^{(1)}_R(\hat\omega,\hat k) + \cdots\nn\\
&=(r_0)^{-2\nu}\left(\fft{\hat\omega}z\right)^{2\nu/z}\left[\fft{\hat B_+}{\hat A_+}
+\kappa\mathcal A_{ij}\hat k^i\hat k^j\fft{zL}{2\nu r_0\hat\omega}
\left(\fft{\psi_+(\hat\omega/z)}{\hat A_+}\right)^2+\cdots\right].
\label{eq:shift1}
\end{align}

In the WKB approximation, (\ref{eq:WKB+}) we find
\begin{equation}
\hat A_+=\sqrt{\fft{z}\nu}e^{-i\pi/4}\lim_{\epsilon\to0}\epsilon^{\nu/z}e^{S(\epsilon,\zeta_*)},\qquad
\hat B_+=\sqrt{\fft{z}\nu}\fft{e^{i\pi/4}}2\lim_{\epsilon\to0}\epsilon^{-\nu/z}e^{-S(\epsilon,\zeta_*)}.
\end{equation}
The retarded Green's function then takes the form
\begin{equation}
G_R(\hat\omega,\hat k)=(r_0)^{-2\nu}\fft12\left(\fft{\hat\omega}z\right)^{2\nu/z}\lim_{\epsilon\to0}\epsilon^{-2\nu/z}
e^{-2S(\epsilon,\zeta_*)}\left[i+\kappa\mathcal A_{ij}\hat k^i\hat k^j\fft{L}{r_0}\fft{\Psi^2(\hat\omega/z)}
{|\nu^2+\hat k^2-\hat\omega^2|^{1/2}}\right],
\label{eq:GRWKB}
\end{equation}
where
\begin{equation}
\Psi(\hat\omega/z)=\begin{cases}
e^{S(\hat\omega/z,\zeta_*)}+\ft{i}2e^{-S(\hat\omega/z,\zeta_*)},&\hat\omega/z<\zeta_*;\\
e^{i\pi/4+i\Phi(\zeta_*,\hat\omega/z)},&\hat\omega/z>\zeta_*.\end{cases}
\label{eq:PsiWKB}
\end{equation}
Recall that $\zeta_*$ is the classical turning point in the WKB potential.

\section{Properties of the perturbed Green's function}
\label{sec:expsup}

We now wish to examine the general features of the retarded Green's function and its ability to probe
bulk perturbations localized at $r_0$.  Since no analytic solution is available for generic values of the critical
exponent $z$, we turn to the WKB  approximation, (\ref{eq:GRWKB}).  Here it is important to recall that
the real part of the unperturbed Green's function, $G_R^{(0)}$, cannot be obtained within the WKB framework,
as the holographic prescription requires the extraction of the normalizable mode from the bulk wavefunction
that is dominated by the non-normalizable mode.  Since the non-normalizable mode can be chosen to be
real, it will not affect the imaginary component of the Green's function.

Taking the imaginary part of (\ref{eq:GRWKB}) then gives the spectral function
\begin{equation}
\chi(\hat\omega,\hat k)=(r_0)^{-2\nu}\left(\fft{\hat\omega}{z}\right)^{2\nu/z}\left(1+\fft{\kappa\mathcal A_{ij}\hat k^i\hat k^jL}{r_0|\nu^2+\hat k^2-\hat\omega^2|^{1/2}}\mathcal F(\hat\omega)\right)\lim_{\epsilon\to0}\epsilon^{-2\nu/z}e^{-2S(\epsilon,\zeta_*)}.
\label{eq:lifspec}
\end{equation}
Here
\begin{equation}
\mathcal F(\hat\omega)=\begin{cases}1,&\hat\omega/z<\zeta_*;\\
\cos 2\Phi(\zeta_*,\hat\omega/z),&\hat\omega/z>\zeta_*.\end{cases}
\end{equation}
is a factor of $\mathcal O(1)$ that captures the oscillatory behavior of the solution when the WKB solution is
in the classically allowed region.  The dominant behavior of the spectral function arises from the tunneling
factor $e^{-2S}$, and is the same for the zeroth and first order terms.  We thus find power-law behavior
at high frequencies and exponential suppression governed by (\ref{eq:chiexp}) at low frequencies.  Moreover,
this suppression of spectral weight is roughly independent of the position $r_0$ of the perturbation.
Heuristically, this can be understood since the location of the perturbation is unimportant in the absence of
any quasiparticles that can probe it.

Although the real part of $G_R^{(0)}$ cannot be determined from the WKB approximation, the real part,
and hence magnitude of $G_R^{(1)}$ can be trusted, as the first order perturbation is parametrically small and
can be isolated from the non-normalizable component of the bulk wavefunction.  Of course, the only 
physical observable is the total Green's function, $G_R=G_R^{(0)}+G_R^{(1)}+\cdots$.  Nevertheless, it is
instructive to examine the magnitude of $G_R^{(1)}$ by itself.  Combining (\ref{eq:GRWKB}) and
(\ref{eq:PsiWKB}) gives
\begin{equation}
|G_R^{(1)}(\hat\omega,\hat k)|\approx(r_0)^{-2\nu}(\kappa\mathcal A_{ij}\hat k^i\hat k^j)\fft{L}{2r_0}
\left(\fft{\hat\omega}z\right)^{2\nu/z}
\lim_{\epsilon\to0}\fft{\epsilon^{-2\nu/z}e^{-2S(\epsilon,\min(\hat\omega/z,\zeta_*))}}{|\nu^2+\hat k^2-\hat\omega^2|^{1/2}},
\label{eq:Gmag}
\end{equation}
In particular, the WKB tunneling factor $e^{-2S}$ suppresses the magnitude of $G_R^{(1)}$ up to the location
of the perturbation, but not beyond the classical turning point $\zeta_*$.  Beyond that, there is no additional
suppression as the wavefunction becomes oscillatory.

\begin{figure}[t]
\begin{center}
\includegraphics[width=10cm]{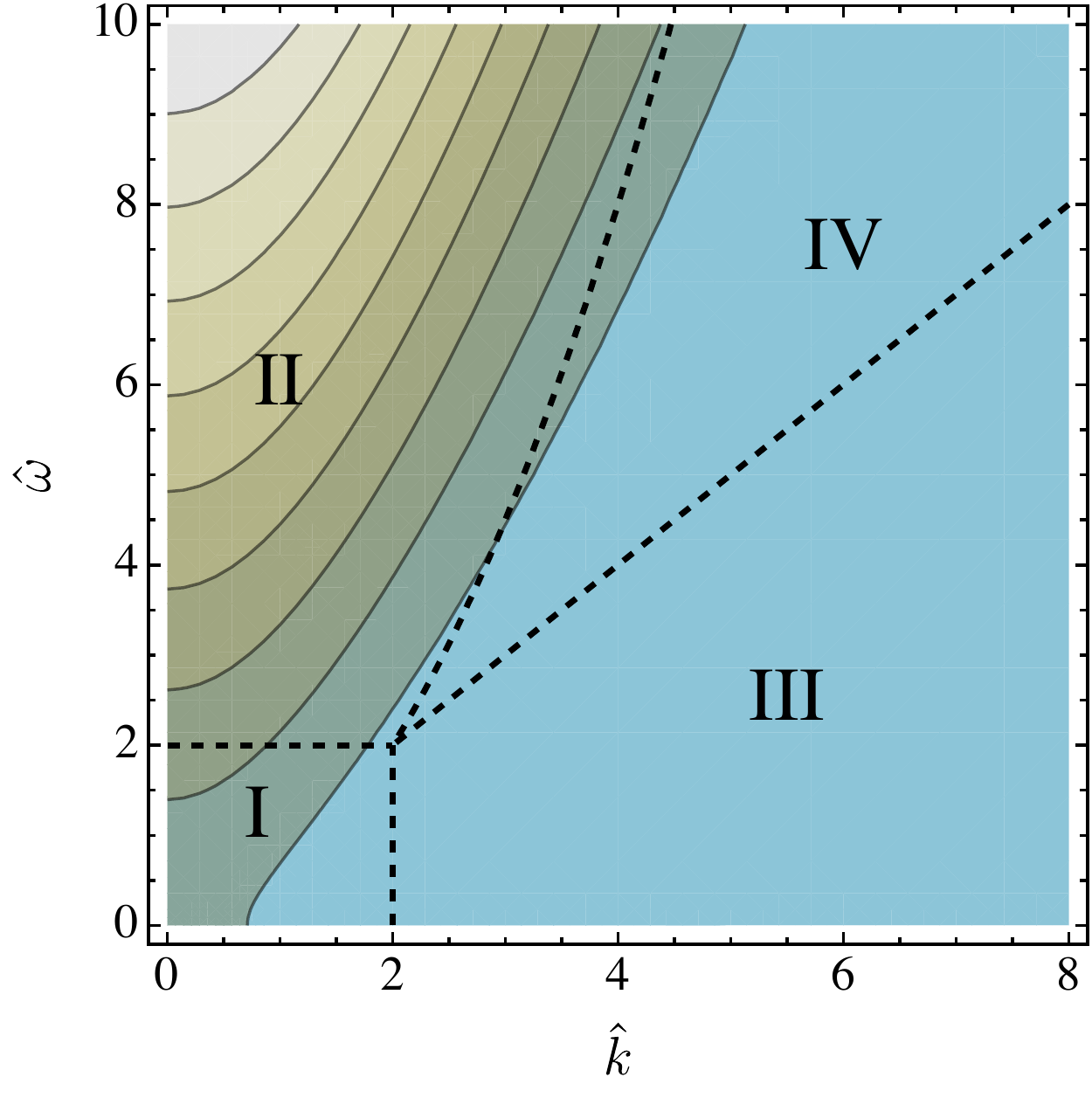}
\end{center}
\caption{Power-law versus exponential suppression in the perturbed Green's function $|G_R^{(1)}(\hat\omega,\hat k)|$.  Regions I and II have power-law behavior, while $|G_R^{(1)}|$ is exponentially suppressed in
regions III and IV.  Here we have taken $z=2$ and $\nu=2$.}
\label{fig:regions}
\end{figure}

In terms of the scaled quantities $\hat\omega$ and $\hat k$, the WKB potential (\ref{eq:pot}) takes the form
\begin{equation}
\hat U(\zeta)=\fft{\nu^2}{(z\zeta)^2}+\fft{\hat k^2/\hat\omega^{2/z}}{(z\zeta)^{2(1-1/z)}}-1.
\label{eq:scapot}
\end{equation}
The competition of the first two power-law terms determines the nature of the WKB wavefunction.  For $(\hat\omega/\nu)
\gg(\hat k/\nu)^z$, the classical turning point lies in the steep $1/\zeta^2$ region, and the WKB result yields power-law
behavior.  However, for $(\hat\omega/\nu)\lesssim(\hat k/\nu)^z$, a tunneling region develops, and the
wavefunction is exponentially suppressed as we approach the horizon.  What this indicates is that $|G_R^{(1)}|$ will have
both power-law behavior and exponential suppression, with the regions roughly demarcated by $(\hat\omega/\nu)
\approx(\hat k/\nu)^z$.  More precisely, we may highlight four cases
\begin{center}
\begin{tabular}{lll}
Case&Region&Behavior\\
\hline
I&$\hat\omega/\nu<1$ and $\hat k/\nu<1$&Power law\\
II&$\hat\omega/\nu>1$ and $\hat\omega/\nu>(\hat k/\nu)^z$&Unsuppressed ($\hat\omega>z\zeta_*$)\\
III&$\hat k/\nu>1$ and $\hat k/\nu>\hat\omega/\nu$&Partial exponential suppression\\
IV&$\hat\omega/\nu>\hat k/\nu$ and $\hat\omega/\nu<(\hat k/\nu)^z$&Maximum exponential suppression
($\hat\omega>z\zeta_*$)
\end{tabular}
\end{center}
Cases II and IV correspond to the perturbation at $r_0$ lying to the right (towards the IR) of the classical turning
point $\zeta_*$.  The behavior of the perturbed Green's function, $|G_R^{(1)}|$, is shown in Fig.~\ref{fig:regions}.

\subsection{The $z=2$ case}

For generic values of the critical exponent $z$, the WKB potential (\ref{eq:scapot}) is non-analytic in $\zeta$.  However,
for $z=2$, it simplifies to $\nu^2/(4\zeta^2)+\hat k^2/(2\hat\omega\zeta)-1$, and the WKB integral (\ref{eq:tunneling})
can be performed exactly, with the result
\begin{align}
S(\epsilon,\zeta)&=-\fft\nu2\log\epsilon+\fft\nu2\left[\sqrt{1+4\zeta(\alpha-\zeta)/\nu^2}-1-\log\left(\fft\alpha{\nu^2}
+\fft1{2\zeta}(1+\sqrt{1+4\zeta(\alpha-\zeta)/\nu^2})\right)\right]\nn\\
&\quad-\fft\alpha2\left[\sin^{-1}\fft{\alpha-2\zeta}{\sqrt{\alpha^2+\nu^2}}-\sin^{-1}\fft\alpha{\sqrt{\alpha^2+\nu^2}}\right],
\label{eq:Swkb}
\end{align}
where we have taken the limit $\epsilon\to0$, and where $\alpha=\hat k^2/2\hat\omega$.  This expression is valid
up to the classical turning point given by $\zeta_*=(\alpha+\sqrt{\alpha^2+\nu^2})/2$.  The divergent factor
$-(\nu/2)\log\epsilon$ cancels the $\epsilon^{-\nu}$ term in (\ref{eq:Gmag}), as it must.

So long as the location of the background perturbation, $r_0$, is to the left of the tunneling region, we do not expect any exponential suppression in the magnitude of $G_R^{(1)}$.  This corresponds to cases I and II above.
For $\hat\omega\ll1$ and $\hat k\ll1$, which essentially corresponds to $r_0$ approaching the boundary, we may expand the WKB integral to obtain
\begin{equation}
|G_R^{(1)}(\hat\omega,\hat k)|\approx(\kappa\mathcal A_{ij}\hat k^i\hat k^j)\fft{L}{2\nu r_0^{2\nu+1}}
\left(1+\fft{\nu+2}{4\nu^2}\hat\omega^2-\fft{\nu+1}{2\nu^2}\hat k^2+\cdots\right).
\label{eq:z=2bdy}
\end{equation}
This remains unsuppressed in the zero frequency limit, provided $\hat k$ is fixed and taken to be small.

We now consider cases III and IV, where we expect to see exponential suppression.  Here the WKB expression (\ref{eq:Swkb}) can be further approximated by neglecting terms that do not contribute toward exponential suppression.  The result gives
\begin{equation}
|G_R^{(1)}(\hat\omega,\hat k)|\approx(\kappa\mathcal A_{ij}\hat k^i\hat k^j)\fft{L}{2r_0\hat k}
\left(\fft{e\hat k}{2\nu r_0}\right)^{2\nu}\fft{e^{-\Xi(u){\hat k^2}/{\hat\omega}}}{\sqrt{|1-u|}},
\end{equation}
where
\begin{equation}
\Xi(u)=\begin{cases}\sqrt{u(1-u)} +\sin^{-1}\sqrt{u},&0\le u\le 1;\\
\pi/2,&\mbox{otherwise}.\end{cases}
\label{eq:Xifunc}
\end{equation}
Here we have defined
\begin{equation}
\label{eq:u}
u =\left(\fft{\hat\omega}{\hat k}\right)^2\qquad\mbox{with}\quad0\lesssim u\lesssim1.
\end{equation}
The function $\Xi(u)$ interpolates between the boundary and turning point, and describes the exponential suppression (or lack thereof) as a function of $u$.  In particular, $\Xi(u)$ begins at zero for $\hat\omega=0$ and ends at $\pi/2$, so that the maximum exponential suppression is of the form $|G_R^{(1)}|\sim e^{-\pi\hat k^2/2\hat\omega}$, in agreement with the corresponding asymptotic expression, (\ref{eq:Gshifts}), in the limit where the location of the perturbation, $r_0$, approaches the horizon.

Of course, the $z=2$ case admits an exact perturbative solution in terms of Whittaker functions.  The zeroth order
expression \eqref{eq:GRz2} can be written in terms of $\hat\omega$ and $\hat k$ as
\begin{equation}
G_R^{(0)}(\hat\omega,\hat k)=\fft{\hat\omega^\nu}{2\pi r_0^{2\nu}}\fft{\Gamma(-\nu)}{\Gamma(\nu)}\left|e^{\pi\alpha/4}\Gamma\left(\fft{1+\nu+i\alpha}2\right)\right|^2(1+e^{-\pi\alpha}e^{-i\pi\nu}),
\end{equation}
where we made use of the reflection formula $\Gamma(z)\Gamma(1-z)=\pi\csc(\pi z)$, and where
$\alpha=\hat k^2/2\hat\omega$.  The low and high frequency expansion of the zeroth order Green's function has the form
\begin{equation}
G_R^{(0)}(\hat\omega,\hat k)\approx\begin{cases}
\left(\fft{\hat k}{2r_0}\right)^{2\nu}\fft{\Gamma(-\nu)}{\Gamma(\nu)}(1+e^{-\pi\hat k^2/2\hat\omega}e^{-i\pi\nu}+\cdots),&\hat\omega\ll\hat k^2;\\
\left(\fft{\hat\omega}{4r_0^2}\right)^\nu\fft{\Gamma(-\nu/2)}{\Gamma(\nu/2)}e^{-i\pi\nu/2}\left(1+\fft{i\pi\hat k^2}{4\hat\omega}\tan(\pi\nu/2)+\cdots\right),&\hat\omega\gg\hat k^2.\end{cases}
\end{equation}
The low-frequency expression shows the characteristic exponential suppression of spectral weight, 
$\chi\sim e^{-\pi\hat k^2/2\hat\omega}$ associated with $z=2$.

The first order shift in the Green's function can be obtained by substituting
(\ref{eq:whit}) into the first order correction (\ref{eq:shift1}), with the result
\begin{equation}
G_R^{(1)}(\hat\omega,\hat k)=(\kappa\mathcal A_{ij}\hat k^i\hat k^j)\fft{L}{2\nu r_0^{2\nu+1}}(-i\hat\omega)^{\nu-1}
\left(\fft{\Gamma(\fft{1+\nu+i\alpha}2)}{\Gamma(\nu)}W_{-i\alpha/2,\nu/2}(-i\hat\omega)\right)^2.
\label{eq:shift}
\end{equation}
Using the series and asymptotic forms of the Whittaker function \cite{NIST:DLMF}
\begin{equation}
W_{\kappa,\mu}(z)\approx\begin{cases}
\fft{\Gamma(2\mu)}{\Gamma(\fft12+\mu-\kappa)}z^{\fft12-\mu}+\mathcal O(z^{\fft32-\Re\mu}),&z\to0;\\
e^{-z/2}z^\kappa(1+\mathcal O(1/z)),&z\to\infty,\end{cases}
\end{equation}
we find
\begin{equation}
G_R^{(1)}(\hat\omega,\hat k)\approx(\kappa\mathcal A_{ij}\hat k^i\hat k^j)\fft{L}{2\nu r_0^{2\nu+1}}\times\begin{cases}
1,&\hat\omega\ll1;\\
(-i\hat\omega)^{\nu-1}\left(e^{\pi\alpha/4}\fft{\Gamma\left(\fft{1+\nu+i\alpha}2\right)}{\Gamma(\nu)}\right)^2
e^{-\pi\alpha}e^{i\hat\omega}\hat\omega^{-i\alpha},&\hat\omega\gg1.
\end{cases}
\label{eq:Gshifts}
\end{equation}
In particular, we see that $G_R^{(1)}$ becomes real and independent of frequency for a metric perturbation
near the boundary (i.e.~as $r_0\to0$).  This matches the boundary expansion of the WKB result (\ref{eq:z=2bdy}) at leading
order in $\nu$, and is consistent with the WKB expansion being in inverse powers of $\nu$.
On the other hand, the behavior for a perturbation near the horizon is
more complicated.  The main feature to observe is the exponential factor $e^{-\pi\alpha}=e^{-\pi k^2/2\omega}$,
which suppresses the sensitivity to the perturbation in the low-frequency limit.  

%
%
%
%
%

\subsection{Suppression factors for arbitrary $z$}
\label{sec:zgen}

Guided by the $z=2$ case examined above, we expect to see the same features of power-law versus exponential
suppression in the sensitivity of the retarded Green's function to bulk metric perturbations for Lifshitz theories with
arbitrary critical exponent.  Since analytical results are not available for $z\ne2$, we will focus on the WKB approximation
in order to gain insight into the general behavior of the Green's function.

As we have noted above, the dominant behavior of the spectral function is given by the unperturbed
result (\ref{eq:chiexp}), which is independent of the location $r_0$ of the perturbation.  On the other hand,
the absolute magnitude of the perturbed Green's function, $|G_R^{(1)}|$, can serve as a probe of $r_0$.  In general,
we expect the WKB result (\ref{eq:Gmag}) to give rise to a partial suppression of the form
\begin{equation}
|G_R^{(1)}(\hat\omega,\hat k)|\approx(\kappa\mathcal A_{ij}\hat k^i\hat k^j)\fft{L}{2r_0\hat k}\left(\fft{e\hat k}{2\nu r_0}\right)^{2\nu}\fft{e^{-\Xi(u)(\hat k^z/\hat\omega)^{\fft1{z-1}}}}{\sqrt{|1-u|}}.
\end{equation}
The function $\Xi(u)$ can be obtained by splitting up the WKB integral based on $\zeta_\star$, the crossover scale where the $\alpha/\zeta^{2(1-1/z)}$ term dominates the $\nu^2/\zeta^2$ term in the potential. Assuming the regions are well-separated, which holds in the small $\hat\omega$ limit, we can perform a matched asymptotic expansion
\begin{equation}
S(\epsilon,\hat\omega/z)\sim\int_\epsilon^{\zeta_*}\sqrt{\fft{\nu^2}{(z\zeta)^2}+\fft{\alpha}{(z\zeta)^{2(1-1/z)}}}d\zeta+\int_{\zeta_*}^{\hat\omega/z}\sqrt{\fft{\alpha}{(z\zeta)^{2(1-1/z)}}-1}d\zeta.
\end{equation}
The first term cancels the power-law divergent prefactor in \eqref{eq:Gmag} and provides an overall power-law factor,
while the second term gives the $r_0$ dependent suppression factor
\begin{equation}
\Xi(u)=2\sqrt{u^{1/(z-1)}(1-u)}\left[1+\fft{z-1}{2z-1}u\;{}_2F_1\left(1,\fft{3z-2}{2(z-1)},\fft{4z-3}{2(z-1)},u\right)\right].
\end{equation}
As in (\ref{eq:u}), the parameter $u=(\hat\omega/\hat k)^2$ goes from zero at the boundary to one at
the classical turning point.  The expression for $\Xi(u)$ reduces to (\ref{eq:Xifunc}) for the case of $z=2$.

\begin{figure}[t]
\begin{center}
\includegraphics{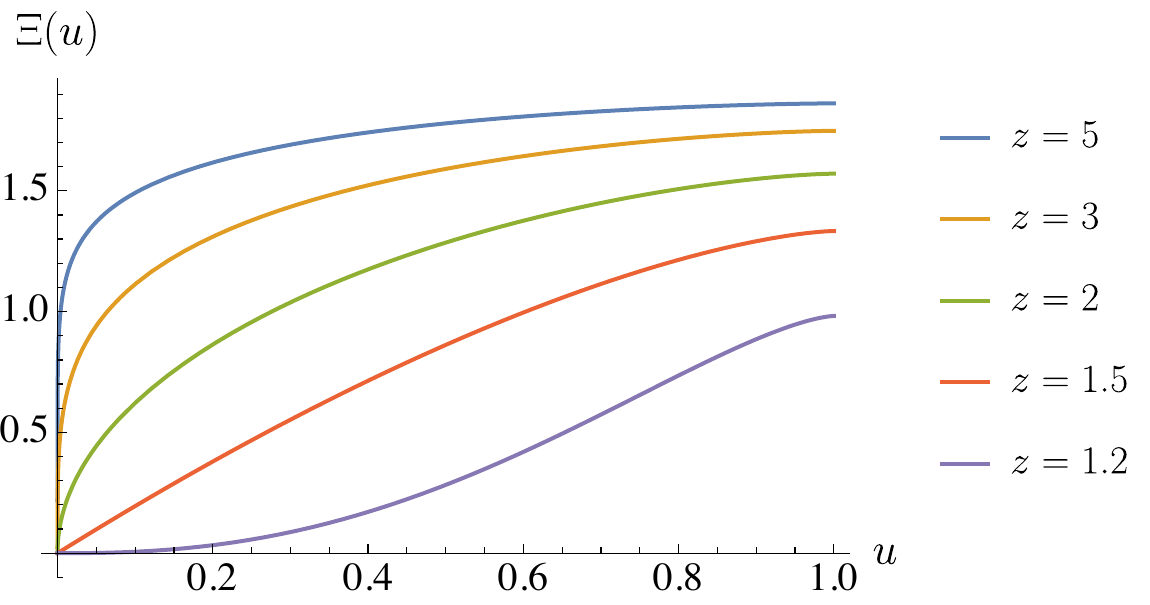}
\end{center}
\caption{The suppression factor $\Xi(u)$ for various Lifshitz critical exponents.  The classical turning point
is located at $u=1$ and the boundary is at $u=0$.  For $u>1$ we have simply $\Xi(u>1)=\Xi(1)$.}
\label{fig:fig2}
\end{figure}

The suppression factor $\Xi(u)$ attains its maximum value at the classical turning point $u=1$.  Taking the
limit gives
\begin{equation}
\Xi(1)=\fft{\sqrt{\pi}}{z}\fft{\Gamma\left(\fft1{2(z-1)}\right)}{\Gamma\left(\fft{z}{2(z-1)}\right)},
\end{equation}
in agreement with (\ref{eq:chiexp}).  The behavior of $\Xi(u)$ for several values of the critical exponent
is shown in Fig.~\ref{fig:fig2}.

\section{Conclusions}
\label{sec:con}

One of the underlying features of holography is the general relation between the bulk radial direction and
energy scale in the dual field theory, with the boundary corresponding to the UV and the deep interior to
the IR.  While this relation can be made more precise in several ways, we have probed the bulk geometry
by studying the effect of a radially localized perturbation on the holographic Green's function.  In particular,
we expect the location of the perturbation, $r_0$, to be encoded through the dependence of
$G_R(\hat\omega,\hat k)$ on the dimensionless frequency $\hat\omega=\omega r_0^z$ and
wavenumber $\hat k=kr_0$.

In the relativistic ($z=1$) case, a straightforward computation gives the spectral function
\begin{equation}
\chi(\omega,k)=\fft{2\pi}{\nu\Gamma(\nu)^2}\left(\fft{q}2\right)^{2\nu}\left(1+\kappa\mathcal{A}_{ij}
\hat k^i\hat k^j\fft{L}{r_0\sqrt{|\nu^2-\hat q^2|}}\mathcal F(\hat q)\right),
\label{eq:z1chi}
\end{equation}
where $q=\sqrt{\omega^2-k^2}$ and $\hat q=qr_0$.  (The spectral function vanishes for $\omega<k$, as
there is no spectral weight `under' the relativistic dispersion relation $\omega=k$.)  The function
$\mathcal F(\hat q)$ arises from the imaginary part of the bulk to boundary propagator, and takes the form
\begin{equation}
\mathcal F(\hat q)=-\sqrt{|\nu^2-\hat q^2|}\pi J_\nu(\hat q)Y_\nu(\hat q)\approx\begin{cases}1,&\hat q\lesssim\nu;\\
\cos(\pi\nu-2\hat q),&\hat q\gtrsim\nu.\end{cases}
\label{eq:Fhat1}
\end{equation}
The inclusion of the square-root factor is somewhat artificial, but is used to bring this expression for the $z=1$
spectral function into the WKB form of (\ref{eq:lifspec}).  The dominant feature that we observe is the power-law behavior
of the spectral weight, $\chi\sim q^{2\nu}$, which is consistent with conformal invariance.  This is broken by the
bulk perturbation at $r_0$, as can be seen through the explicit dependence on $r_0$ and $\hat q$.

We are now in a position to address how well we see the perturbation at location $r_0$.  Since $\mathcal F(\hat q)$ is a function of $\mathcal O(1)$ throughout its domain, the imprint of the perturbation on the spectral function
is of the form
\begin{equation}
\fft{\delta\chi}{\chi}\sim \fft{k^2r_0}{\sqrt{|\nu^2-(qr_0)^2|}}.
\label{eq:dcc1}
\end{equation}
At fixed wavenumber $k$, we see that $\delta\chi/\chi$ approaches a constant, independent of frequency,
for small $q$, but falls off as $k^2/q$ for large $q$.  The transition point is given by $q\approx\nu/r_0$.  Moreover, the large $q$ behavior is oscillatory, as can be seen from (\ref{eq:Fhat1}).  From a holographic
point of view, we expect high frequency probes to be sensitive to the UV (i.e.~the region close to the boundary), and this is consistent with the
behavior seen here.  In particular, for fixed $r_0$, $\delta\chi/\chi$ remains mostly constant up until $\omega^2\approx
k^2+\nu^2/r_0^2$, and then begins to fall off, eventually reaching a high-frequency behavior
$\delta\chi/\chi\sim 1/\omega$.

Of course, in order to actually see the perturbation, there must be sufficient spectral weight at the frequency of the probe.  In the $z=1$ case, this shows up as the power-law prefactor $\chi\sim q^{2\nu}$ in (\ref{eq:z1chi}).
However, this behavior is rather different in the Lifshitz case.  In particular, the WKB result (\ref{eq:lifspec})
gives the leading order behavior $\chi\sim e^{-2S}$ along with the perturbation
\begin{equation}
\fft{\delta\chi}\chi\sim\fft{k^2r_0}{\sqrt{|\nu^2+(kr_0)^2-(\omega r_0^z)^2|}},
\end{equation}
which is a natural generalization of (\ref{eq:dcc1}) to the Lifshitz case.  Again, for fixed wavenumber and $r_0$, this is roughly constant at small frequencies and falls off as $1/\omega$ at high frequencies.
However, the visibility of this perturbation crucially depends on whether this is sufficient spectral weight at
the frequency of the probe.  At sufficiently large frequencies, there is no suppression, and in principle the
entire bulk geometry is visible.  On the other hand, at low frequencies, the spectral weight is exponentially
suppressed, and no probes of the bulk geometry are available.  Taken together with $\delta\chi/\chi$, we see that bulk Lifshitz perturbations in the IR can be partially hidden, as only high frequency probes are available, and such probes are power-law suppressed deep in the bulk.

The exponential suppression of spectral weight below the non-relativistic dispersion relation is a
universal feature of Lifshitz spacetimes, and furthermore provides an obstruction to the construction
of a smearing function for the bulk reconstruction of local boundary operators \cite{Keeler:2013msa}.
The lack of a smearing function suggests that the bulk geometry itself cannot be fully recovered from
the boundary, and this is consistent with the exponentially suppressed sensitivity to bulk perturbations
in either the small $\omega$ or large $k$ limit.

Finally, although we have focused on spatially homogeneous bulk perturbations, it would be natural to
consider the case of broken translational symmetry, such as would occur with a bulk lattice.  While
such lattices typically extend to the boundary, the interesting physical regime is often at long
wavelengths and low frequencies.  In this case, the bulk geometry near the horizon is relevant, and as
we have seen there is partial suppression of its imprint on the holographic Green's function.  More generally,
we may expect regimes of suppressed spectral weight any time relativistic invariance is broken, and this
can lead to interesting implications for holographic systems.

\acknowledgments

The initial idea of exploring perturbations to the Lifshitz bulk arose out of discussions with C.\ Keeler and
G.\ Knodel. PVR thanks Joshua Foster and Weishun Zhong for helpful discussions.
This work was supported in part by the US Department of Energy under Grant No.~DE-SC0007859.

\bibliographystyle{JHEP}
\bibliography{lifshitz-refs}

\end{document}